\documentclass[twocolumn,showpacs,superscriptaddress,amsmath,amssymb]{revtex4}
\usepackage{graphicx}
\usepackage{dcolumn}
\usepackage{bm}
\usepackage{bm}
\def\bs{\boldsymbol}

\def\sech{{\rm sech}\aa}          \def\csch{{\rm csch}\aa}
\def\Arg{{\rm Arg}\aa}            \def\Arctg{{\rm Arctg}\aa}
\def\arctg{{\rm arctg}\aa}        \def\vs{\vskip}
\def\tg{{\rm tg}\aa}              \def\ctg{{\rm ctg}\aa}
\def\sh{{\rm sh}\aa}              \def\ch{{\rm ch}\aa}
\def\th{{\rm th}\aa}              \def\tan{{\rm tan}\aa}
\def\cth{{\rm cth}\aa}            \def\f{\left}  \def\g{\right}

\def\grad{{\rm grad\hskip3pt}}    \def\div{{\rm div\hskip3pt}}
\def\huaD{\mathcal{D}}            \def\huaL{\mathcal{L}}
\def\Re{{\bf Re}\aa}              \def\Im{{\bf Im}\aa}

\def\etc{{\it etc.}}              \def\ie{{\it i.e. }}
\def\cf{{\it cf.}}                \def\eg{{\it e.g. }}
\def\cH{{\cal H}}                 \def\cV{{\cal V}}

\def\pt{p_{\rm T}}                \def\mt{m_{\rm T}}
\def\bd{\begin{document}}   \def\ed{ \end{document} }
\def\ie{{\it i.e.}\ }   \def\eg{{\it e.g.}\ }   \def\cf{{\it cf.}\ }
\def\etc{{\it etc.}\ }  \def\P{{\rm P}}
\def\bs{\boldsymbol} \def\nd{\noindent} \def\nbf{\nd\bf} \def\mn{\vskip0.5cm\nd}
\def\aa{\hskip3pt} \def\aaa{\hskip1.5pt}  \def\np{\newpage}
\def\md{\vskip0.3cm}
\def\bec{\begin{center}}    \def\eec{\end{center}}
\def\bct{\begin{center}}    \def\ect{\end{center}}  \def\cl{\centerline}
\def\hs{\hskip}  \def\vs{\vskip}  \def\lrw{\longrightarrow}
\def\bmp{\begin{minipage}}    \def\emp{\end{minipage}}
\def\beq{\begin{equation}}    \def\eeq{\end{equation}}
\def\bea{\begin{eqnarray}}    \def\eea{\end{eqnarray}}
\def\bes{\begin{eqnarray*}}    \def\ees{\end{eqnarray*}}
\def\bpm{\begin{pmatrix}} \def\epm{\end{pmatrix}}
\def\ben{\begin{enumerate}} \def\een{\end{enumerate}}
\def\btb{\begin{tabular}} \def\etb{\end{tabular}}
\def\btbb{\begin{tabbing}} \def\etbb{\end{tabbing}}
\def\af{\alpha} \def\bt{\beta}  \def\gm{\gamma}  \def\tr{{\rm tr}\,}
\def\lm{\lambda}  \def\Lm{\Lambda} \def\spic{^{\footnotesize(\rm S)}}
\def\hpic{^{\footnotesize(\rm H)}}   \def\upic{^{\footnotesize(\rm U)}}
\def\ipic{^{\footnotesize(\rm I)}}  \def\hn{\hat{\bs n}}
\def\sech{{\rm sech}\,}  \def\Arg{{\rm Arg}\,} \def\Arctg{{\rm Arctg}\,}
\def\arctg{{\rm arctg}\,}  \def\nbr{\nonumber} \def\dt{\delta}
 \def\Dt{\Delta} \def\ep{\epsilon} \def\ve{\varepsilon}
 \def\sm{\sigma}   \def\Sm{\Sigma}      \def\ta{\theta}  \def\Ta{\Theta}
 \def\om{\omega}   \def\Om{\Omega}    \def\kp{\kappa}  \def\gm{\gamma}
\def\tan{{\rm tan}\,}  \def\vf{\varphi}  \def\vt{\vartheta}
 \def\cH{{\cal H}}   \def\cV{{\cal V}}   \def\cD{{\cal D}\,}
 \def\cK{{\cal K}\,}   \def\Uvf{U_{\bs \vf}}
 \def\Gm{\Gamma}    \def\ih{\frac{i}{\hbar}}  \def\cI{{\cal I}}
 \def\tg{{\rm tg}\,}      \def\ctg{{\rm ctg}\,} \def\csch{{\rm csch}\,}
\def\sh{{\rm sh}\,}      \def\ch{{\rm ch}\,}   \def\th{{\rm th}\,}
\def\cth{{\rm cth}\,} \def\C{{\rm C}}   \def\grad{{\rm grad\hskip3pt}}
\def\div{{\rm div\hskip3pt}}  \def\L{{\rm L\hskip3pt}}
\def\lra{\longrightarrow}  \def\bone{{\bf 1}}
\def\qq{\qquad}   \def\fc{\frac}   \def\fnsz{\footnotesize}
\def\inint{\int_{-\infty}^{\infty}}  \def\ol{\overline}
 \def\ben{\begin{enumerate}} \def\een{\end{enumerate}}
\def\qd{\quad}  \def\qqd{\qquad} \def\btm{\begin{itemize}}
\def\etm{\end{itemize}}   \def\pl{\partial}  \def\huaN{\mathcal{N}}
\def\huaD{\mathcal{D}}  \def\huaL{\mathcal{L}} \def\d{{\rm d}}
\def\ddz{{\rm d}\over{{\rm d}z}}  \def\dv{{\rm d}} \def\huaG{\mathcal{G}}
\def\Re{{\bf Re}\;} \def\Im{{\bf Im}\;}  \def\g{\right}  \def\e{{\rm e}}
\def\f{\left} \def\r{\right} \def\la{\langle} \def\ra{\rangle}
\def\npg{\newpage} \def\ihb{\frac{i}{\hbar}}
\def\dbar{{\rm d}\hskip-5.6pt \rule[1.8mm]{2.0mm}{0.18mm}\hskip2pt}
\def\sdbar{{\rm d}\hskip-4.2pt \rule[1.45mm]{1.4mm}{0.12mm}\hskip2pt}
\def\ointcw{\mathop{\int\mkern-21.mu \circlearrowright}} 
\def\ointacw{\mathop{\int\mkern-20.mu \circlearrowleft}} 
\def\sointcw{\mathop{\int\mkern-19.mu \circlearrowright}} 
\def\sointacw{\mathop{\int\mkern-18.mu \circlearrowleft}} 
\def\ssointcw{\mathop{\int\mkern-18.5mu \circlearrowright}} 
\def\ssointacw{\mathop{\int\mkern-17.5mu \circlearrowleft}} 
\def\Solution{\vskip1mm\noindent {$\bs S\bs o\bs l\bs u\bs t\bs i\bs o\bs n$.}\quad}
\def\Res{{\bf Res}} \def\axiom{{\vskip2mm\noindent \bf Axiom\ }}
\def\bcs{\begin{cases}}  \def\ecs{\end{cases}}
\newcommand{\oiint}{\mathop{\makebox[-0,32em][l]
{$\bigcirc$}\int\!\!\!\!\!\int\makebox[-0.5em]{}}}

\begin{document}
\normalsize
\title{Effect of equilibrium phase transition on multiphase transport\\ in relativistic heavy ion collisions
\footnote{supported by NSFC under project 10375025 and by the
Cultivation Fund of the Key Scientific and Technical Innovation
Project£¬Ministry of Education of China NO CFKSTIP-704035.}}

\author{Yu Meiling\footnote{Email:\ yuml@iopp.ccnu.edu.cn} \quad Du Jiaxin
\quad Liu Lianshou\footnote{Email:\ liuls@iopp.ccnu.edu.cn}}

\affiliation{Institute of Particle Physics, Huazhong Normal
University, Wuhan 430079, China}

\begin{abstract}
The hadronization scheme for parton transport in relativistic
heavy ion collisions is considered in detail. It is pointed out
that the traditional scheme for particles being freezed out one by
one leads to serious problem on unreasonable long lifetime for
partons. A super-cooling of the parton system followed by a
collective phase transition is implemented in a simple way. It
turns out that the modified model with a global phase transition
is able to reproduce the experimental longitudinal distributions
of final state particles better than the original one does. The
encouraging results indicate that a relevant parton transport
model for relativistic heavy ion collision should take equilibrium
phase transition into proper account.\end{abstract}



\maketitle

\section{Introduction}

The quark-gluon plasma (QGP) is expected to be formed in heavy-ion
collisions at the Relativistic Heavy-Ion Collider (RHIC). So far
very interesting experimental data have been
collected\;\cite{whitepapers}. There are strong evidences on the
deconfinement of QCD vacuum and the appearance of a (locally)
thermalized partonic system at the early stage of collision. Both
theoretical and experimental investigations are stimulated on the
evolution of the partonic system and the way it transforms to
final state particles.

\begin{figure}[!ht]
\begin{center}
\includegraphics[width=0.5\linewidth]{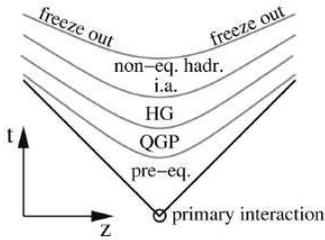}
\caption{\label{stage} The different stages of heavy-ion collisions.}
\end{center}
\end{figure}

The reaction process of relativistic heavy ion collision can be
divided into several stages (see figure~\ref{stage}).
At first the primary interaction creates many partons. These
partons are certainly far from equilibrium and cascade through
interaction among each other. If the colliding nuclei is heavy
enough and the incident energy is high enough the partonic system
is expected to reach equilibrium or local equilibrium, forming
quark-gluon plasma. The system then expands and the temperature
decreases, passing via a phase transition or crossover into the
hadron gas stage. Hadrons continue to cascade until they freeze
out from the collision region, forming final state particles.

At present there is no unique theory that can describe the
reaction process as a whole. Different theoretical approaches are
applied to different stages\;\cite{qgpprocess}. The primary
interaction, creating many partons is often described by
eikonalized parton model\;\cite{hijing}, Gribov-Regge
theory\;\cite{epol} or parton saturation model\;\cite{larry}.
Parton and/or hadron cascade is referred to as the solution of
non-equilibrium transport equation~\cite{transport} and is usually
realized by Monte Carlo
models\;\cite{rqmd}\;\cite{urqmd}\;\cite{ampt_sum}. The
hydrodynamics\;\cite{hydro} or thermal models\;\cite{thermal} are
employed to deal with the parton and/or hadron system in local or
global equilibrium.

Usually, the hydrodynamics or thermal models take (local)
equilibrium as model assumption and do not answer the question on
how the system arrive at (local) equilibrium. The global
properties of the system are the main issues considered in the
model, while the detailed evolution of the constituents ------
partons and/or hadrons is not taken into account.

On the contrary, the transport models follow the evolution of the
partons and/or hadrons in detail through considering their
interaction ------ elastic and/or inelastic scattering, or
cascade. Usually, the model allows the cascade continues to go on
until the interaction ceases and then the parton or hadron freeze
out from the system. A noticeable common property of such model is
that, freeze out of parton or hadron occurs {\it particle by
particle} . Each particle has its own freeze-out time. Such an
approach is acceptable for a hadronic system, where hadrons after
ceasing to interact with other hadrons fly away from the system
freely toward the detectors, but will cause serious problem for a
partonic system.

Partonic and hadronic systems are of different phases. In general,
{\it a phase is a portion of a system that is uniform and has a
definite boundary}\;\cite{kittel}. In particular, the partonic and
hadronic phases exist in different vacua ------ the former is in
perturbation QCD vacuum and the latter in physical vacuum.
Therefore, the transition between partonic and hadronic phases
should be a collective, thermal-equilibrium phenomenon,
accompanied by a vacuum transform, which could not be realized in
a particle-by-particle way.

In the transport models presently on market, the hadronization is
realized parton-wise instead of collectively, and therefore, in
these models there is only {\it hadronization} but no partonic to
hadronic {\it phase transition} in the strict sense. To let a few
partons live for a very long time is inconsistent with the general
believe that the existing time of QGP is about 1-5
fm/$c$~\cite{plasmatime} and hadron freeze out at about 20-40
fm/$c$~\cite{hadrontime}. Even more seriously, when most of the
partons have already hadronized, the system is dominated by
hadrons and the corresponding vacuum is a physical one instead of
a QCD perturbation one. To let some partons survive and fly freely
in such a circumstance is highly unphysical.

The aim of the present paper is to discuss this problem in detail.
We will take as example a presently available model, that has
parton transport implemented. The temperatures of the system at
different time will be extracted using thermal-equilibrium
transverse mass distribution. The partonic system will be allowed
to somewhat {\it super-cooled}, \ie the temperature is allowed to
decrease to lower than the expected phase-transition temperature.
At a certain point all the partons remaining in the system are
forced to coalesce, forming hadrons. Then the hadrons start to
cascade toward freeze out. Thus the difficulty of {\it long-life
parton} is overcome in a simple manner. The phenomenological
consequences of such an approach will be presented and compared
with existing experimental data. The possible reason for the
improvement of the present approach with global hadronization in
comparison with the original model with long-life partons will be
discussed.

The layout of the paper is as the following. A short introduction
to the AMPT model, used as an example for models with parton
transport is given in Section II. A detailed analysis about the
parton and hadron time evolution in AMPT is then presented in
Section III. A super-cooling of the parton system followed by a
collective hadronization scheme is proposed in Section IV together
with the phenomenological consequences on the final hadron
distribution and elliptic flow. Section V is conclusion and
discussion.

\section{\label{section2} A brief introduction to AMPT}

The AMPT model\;\cite{ampt_sum} is based on non-equilibrium
transport dynamics. It contains four main components: the initial
conditions, partonic interactions, conversion from the partonic to
the hadronic matter and hadronic interactions. The initial
conditions, which includes the spacial and momentum distributions
of minijet partons from hard processes and strings from soft
processes, are obtained from the HIJING model in which eikonized
parton model is employed. The time evolution of partons is then
modeled by the ZPC~\cite{zpc} parton cascade model. At present
this model includes only parton-parton elastic scattering with
cross section \beq \sigma_p\simeq
\frac{9\pi\alpha_s^2}{2\mu^2},\eeq where the screening mass $\mu$
is taken to be an input parameter of the model to obtain the
desired total cross section. Two partons will undergo scattering
when the closest distance between them is smaller than
$\sqrt{\sigma/\pi}$.

There are two versions of AMPT model. In the default AMPT, after
ceasing interactions minijet partons are combined with their
parent strings to form excited strings, which are then converted
to hadrons according to the Lund string fragmentation model. While
in the AMPT with string melting, the strings in the initial
conditions are melt to partons first and then interactions among
all the partons are again modeled by ZPC. After partons stop
interacting, a simple quark coalescence model is used to combine
the two nearest partons into a meson and three nearest quarks
(antiquarks) into a baryon (antibaryon). Scatterings among the
resulting hadrons are described by a relativistic transport (ART)
model~\cite{art} which includes baryon-baryon, baryon-meson and
meson-meson elastic and inelastic scatterings.

It turns out that the default AMPT (v1.11) is able to give a
reasonable description on hadron rapidity distributions and
transverse momentum spectra observed in heavy ion collisions at
both SPS and RHIC. However, it fails to reproduce the experimental
data about elliptic flow and two-pion correlation function. On the
other hand, the AMPT model with string melting (v2.11) can well
describe the elliptic flow and two-pion correlation
function~\cite{ampt_flow}\cite{ampt_hbt} but agree baddly with the
hadron rapidity and transverse momentum spectra.

In the following we will utilize the AMPT with string melting
------ AMPT v2.11 to generate Au-Au central collision events
at $\sqrt{s_{\rm NN}}=200$ GeV.
The impact parameter is in the range $b\leq 3$ fm and
the parton cross section is taken to be 10 mb.

\section{\label{section3} The time evolution of partons and hadrons in AMPT}

In AMPT v2.11, each initial parton has a formation time given by
$t_f=E_H/m_{T.H}^2$ with $E_H, m_{T,H}$ the energy and transverse
mass of its parent hadrons. After this formation time the partons
start to scatter with each other and when a parton no longer
scatters with any other parton, it will hadronize. Thus each
parton has its own hadronization time.

\begin{figure}[!ht]
\begin{center}
\includegraphics[width=0.7\linewidth]{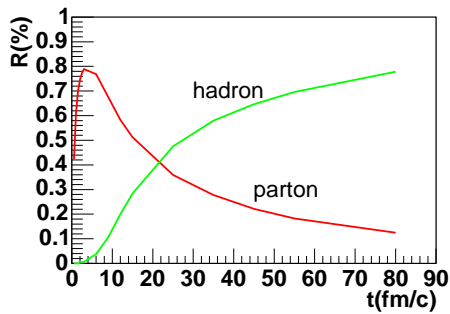}
\caption{\label{phn} The percentage of partons and hadrons,
respectively, in AMPT v2.11 for $\sqrt{s_{NN}}=200$ GeV Au-Au
central collisions with $b\leq 3$ fm and parton cross section 10
mb.}
\end{center}
\end{figure}

In Fig.~\ref{phn} are shown the percentages of parton and hadron,
respectively, at different time after the collision. It can be
seen from the figure that a few hadrons (about 4\%) have already
emerged at $t<5$ fm/$c$. At this time, partons dominate, and
the system as a whole is in the deconfined phase, located in the
perturbation QCD vacuum, with a few hadrons vaporized out.

As time increases, the number of partons decreases while that of
hadrons increases. In this process a parton transforms to hadron
when and only when it ceases to interact with other partons. Thus
a part of parton survive up to an unreasonable long time, e.g.
$t\sim 100$ fm/$c$. This is the difficulty common for this kind of
models mentioned in the Introduction.

For simplicity, we assume that the system arrives at equilibrium
{\it globally} \, instead of locally and the equilibrium is
examined at different {\it time} $t$ instead of at different
intrinsic time $\tau$. Under this assumption the temperature of
the system at different time is determined by thermodynamics.

\begin{figure}[!ht]
\begin{center}
\includegraphics[width=0.45\linewidth]{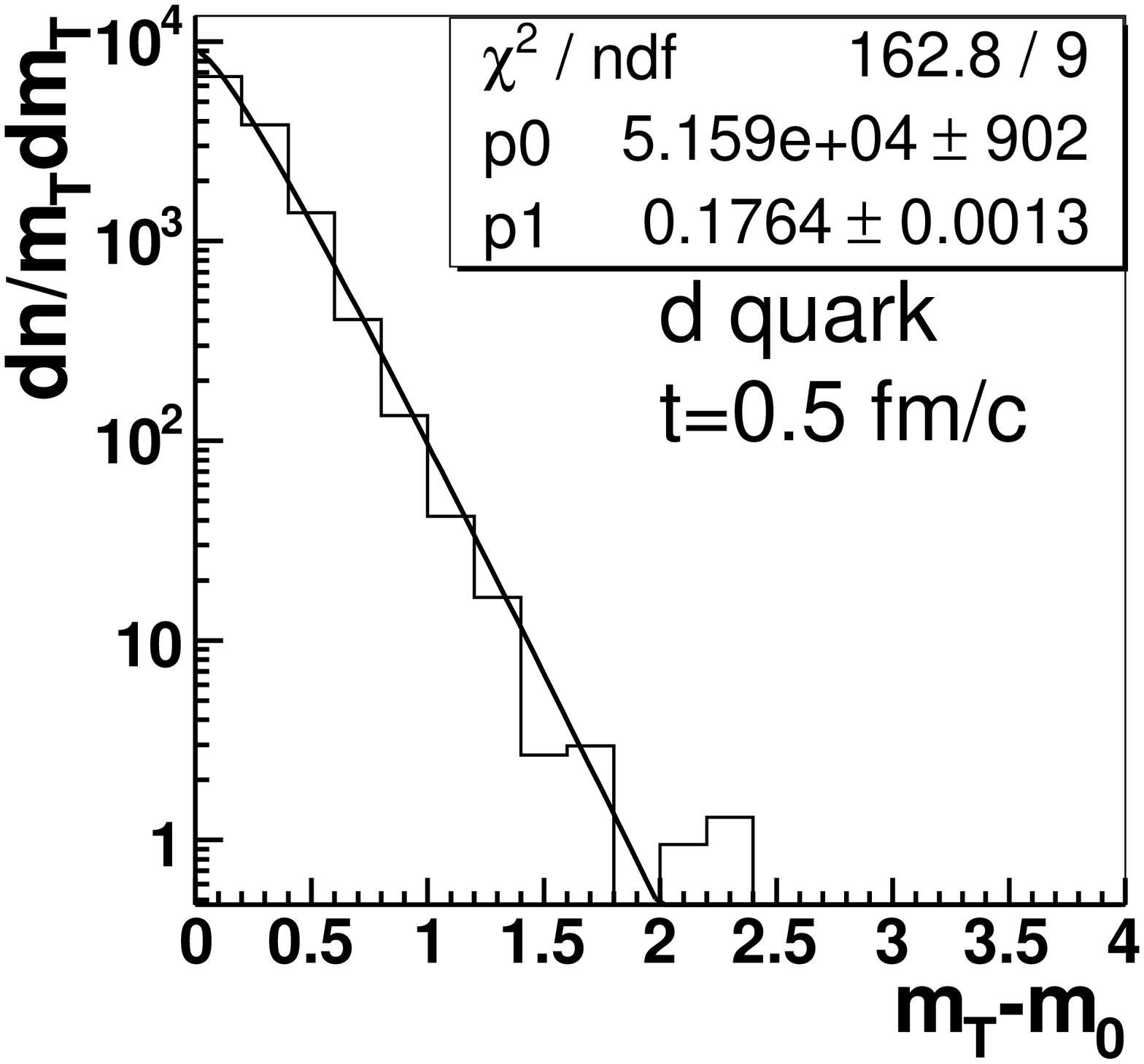}
\includegraphics[width=0.45\linewidth]{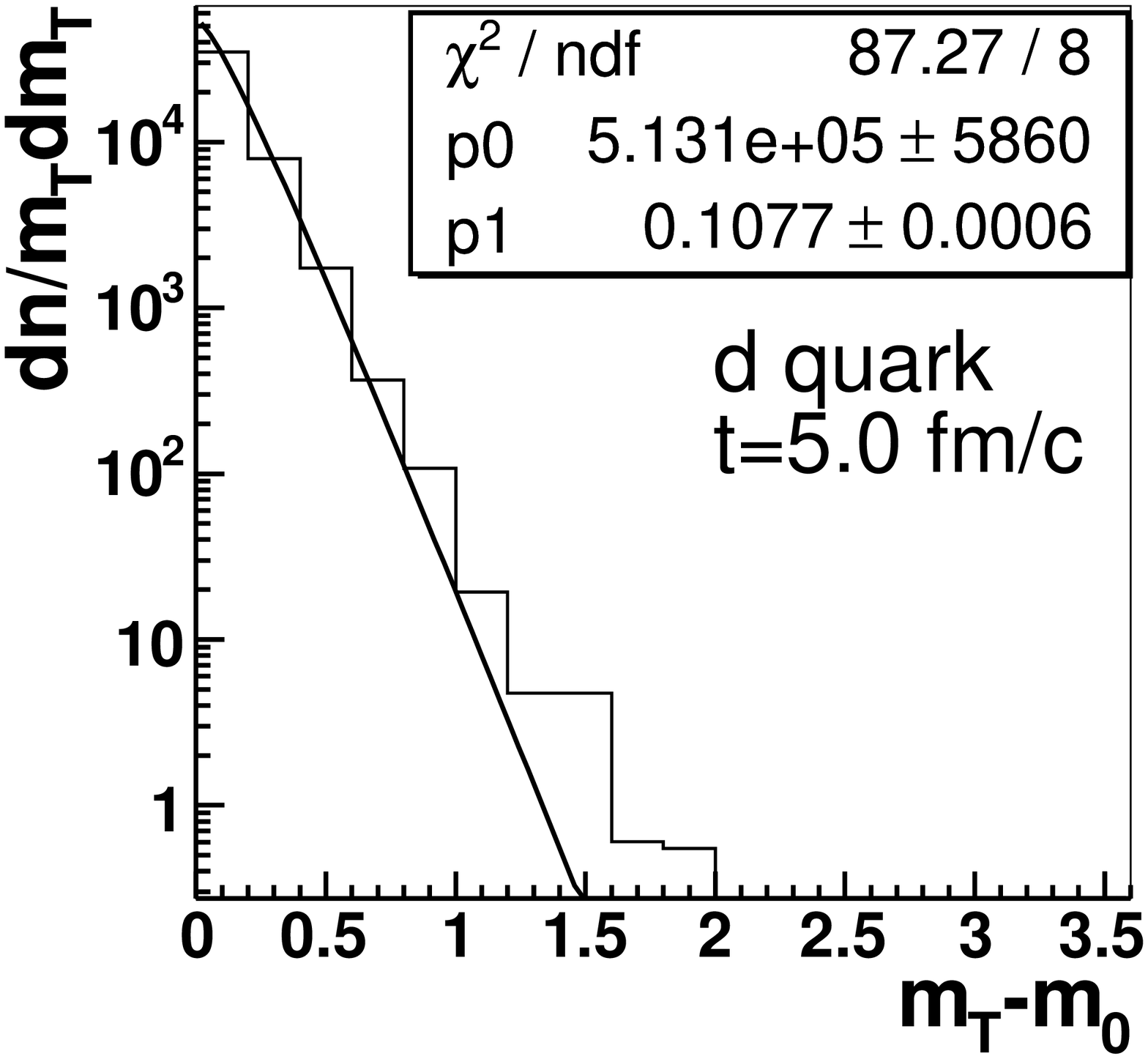}
\caption{\label{phTt} The transverse mass distribution
(histograms) for d-quarks at two different time $t=0.5$ and $5$
fm/$c$ . The lines are the fit to Eq.\;(3)}
\end{center}
\end{figure}

Suppose the expansion of system is a slow and quasi-static
process, then at each time step the system can be regarded as an
equilibrium thermal system with temperature $T$. The invariant
momentum distribution of particles emitted from the thermal system
is~\cite{thermal2} \beq E\frac{d^3n}{d^3p}=\frac{dn}{{dy\mt d\mt
d\phi}}=\frac{gV}{{(2\pi)^3}}Ee^{-(E-\mu)/T},\eeq where $g$ is the
particle spin-isospin degeneracy factor, $\mu$ is the chemical
potential, $V$ is the system volume, $E$ is the energy of the
particle, $y$ is rapidity and $\mt=\sqrt{m^2+\pt^2}$ is the
transverse mass. Integrating with respect to $y$ and $\phi$, we
get the transverse mass distribution \beq \frac{dn}{{\mt
d\mt}}=\frac{gV}{{2\pi^2}}\mt K_1\f(\frac{\mt}{T}\g), \eeq in
which $K_1$ is the hyperbolic Bessel function of rank one. By
fitting the transverse mass distribution, the temperature of the
system can be extracted. As example, in Fig's.\;\ref{phTt} are
shown the transverse mass distributions for d quarks at two
different time $t=0.5$ and $5$ fm/$c$ and the fit to Eq.\;(3). In
the present work we assume the system to be static, and omit the
effect of radial flow on the fitted temperature.

It can be seen from Fig's.\;\ref{phTt} that the temperature of the
parton system is about 180 MeV at $t=0.5$ fm/c, slightly above the
predicted critical temperature $T_{\rm c}=170$ MeV~\cite{lqcd},
while at $t=5$ fm/c the temperature has already arrived at about
110 MeV, \ie in the first few fm/c the temperature decreases
rapidly and soon becomes lower than the expected phase transition
temperature.

\begin{figure}[!ht]
\begin{center}
\includegraphics[width=0.7\linewidth]{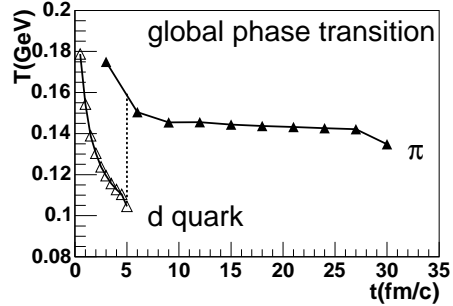}
\caption{\label{Tevolution} The temperature evolution extracted
from parton and hadron transverse mass spectra in AMPT v2.11 with
a collective phase transition implemented at $t=5$\;fm/$c$ for
$\sqrt{s_{NN}}=200$\;GeV Au-Au central collisions with $b\leq 3$
fm and parton cross section 10 mb.}
\end{center}
\end{figure}

\section{\label{section4} A hadronization scheme with a
super-cooling followed by a collective hadronization}

We regard the temperature of parton system decreasing rapidly to
lower than the critical temperature as a {\it supercooling
effect}, i.e. after the formation of QGP, the system lowers its
temperature by expanding and evaporating hadrons, arriving at a
temperature lower than the phase transition temperature. Then at a
certain point all the left partons are coalesced to hadrons. The
duration of the supercooling state is taken as a model parameter
and in the present work we take $t=5$ fm/$c$ (the corresponding
parton temperature is about $110$ MeV) to be the beginning of the
phase transition. At this time, all the remaining partons stop to
interact and start to hadronize. The temperature of the resulting
hadron system is higher than that of the parton system due to the
release of latent heat, and then the temperature decreases again
through expansion, \cf  Fig.\ref{Tevolution}.

Thus we have implemented an equilibrium phase transition with
supercooling to the transport model in a very simple way. Our
purpose is to see how the phase transition affect the final state
hadron distributions.

\begin{figure}[!ht]
\begin{center}
\includegraphics[width=\linewidth]{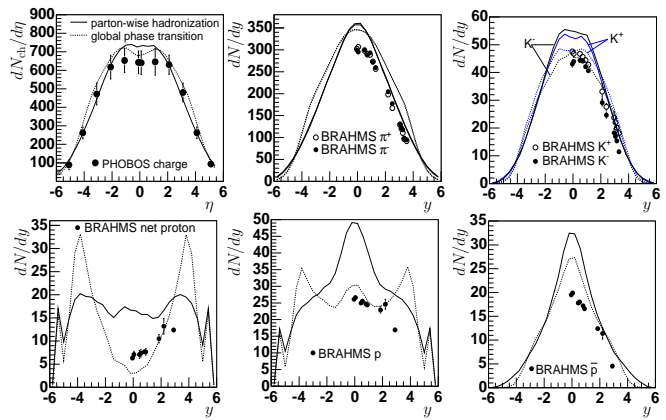}
\caption{\label{rapidity} The rapidity distribution for
$\sqrt{s_{NN}}=200$ GeV Au-Au central collisions. The solid lines
are AMPT v2.11 with parton-wise hadronization and the dashed lines
are that with collective phase transition implemented. The impact
parameter is $b\leq 3$ and parton cross section 10 mb. The dots
are data from PHOBOS 6\% and BRAHMS 5\% central collisions. }
\end{center}
\end{figure}

The rapidity distribution is \beq \frac{\d N}{\d
y}=\frac{1}{\huaN_{\rm ev}}\frac{\Delta n}{\Delta y},\eeq where
$\huaN_{\rm ev}$ is the number of events, $\Delta y$ is the width
of rapidity bin, $\Delta n$ is the number of particles inside the
rapidity bin. In Fig's.~\ref{rapidity} are shown the rapidity
distributions for charged particles, pions, kaons, net-protons,
protons and anti-protons. The solid lines represent the results of
AMPT v2.11 with parton-wise hadronization and the dotted lines are
that with collective phase transition implemented in the
above-mentioned way. Full circles are PHOBOS 0-6\% centrality data
and BRAHMS 0-5\% centrality data. The data of $dN_{\rm ch}/d\eta$
are with both statistical and systematic errors and the other data
are with only statistical ones. The figures show that the rapidity
distributions of the model with phase transition implemented have
a better description of the experimental data, especially on the
kaon and proton rapidity distributions.
As about the transverse distributions we found that they are
almost unaffected by the implementation of global phase transition

The success of AMPT with string melting is that it is able to
describe the elliptic flow data very well through adjusting the
parton cross section. So, whether the implementation of a global
phase transition will destroy this agreement between model and
data is a natural question.

\begin{figure}[!ht]
\begin{center}
\includegraphics[width=0.7\linewidth]{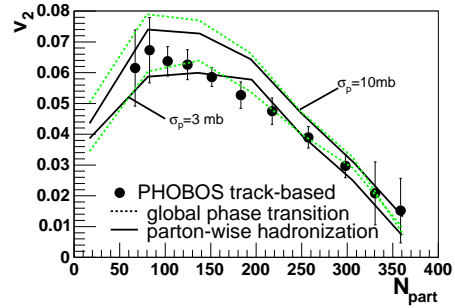}
\caption{\label{flow} Centrality dependence of charged hadron
elliptic flow for $\sqrt{s_{NN}}=200$ GeV Au-Au central
collisions. The solid lines are the AMPT v2.11 with parton-wise
hadronization and dashed lines are that with global phase
transition implemented. The dots are data from PHOBOS
experiments~\cite{v2_exp}.
The impact parameter $b\leq 3$ and parton
cross section is 10mb and 3mb, respectively.}
\end{center}
\end{figure}

In Fig.~\ref{flow} are shown the elliptic flows $v_2$ as a
function of the number of participants calculated from the AMPT
(with string melting) model with parton-wise hadronization and
with collective phase transition, respectively. The parton cross
sections are taken to be 3 mb and 10 mb. It is clear from the
figure that the $v_2$ with collective phase transition implemented
has almost the same shape as that with parton-wise hadronization.
The implementation of global phase transition affects the elliptic
flow little and preserves the agreement between model and
experimental data.

\section{Conclusion and discussion}

In the traditional transport models the particles in the system
are freezed out one by one, each particle has its own freeze-out
time. Extending such an approach to parton transport leads to
serious problem on unreasonable long lifetime for partons.

In order to avoid this problem, we assume the system to have
reached global equilibrium and extract temperature from the system
by thermodynamics formula. As the decreasing of temperature the
parton phase is allowed to hadronize as a whole after a
supercooling stage. It turns out that the modified model with a
global phase transition inherits the success of the original one
in elliptic flow and is able to reproduce the experimental
longitudinal distributions of final state particles better than
the original one does.

\begin{figure}[!ht]
\begin{center}
\includegraphics[width=0.7\linewidth]{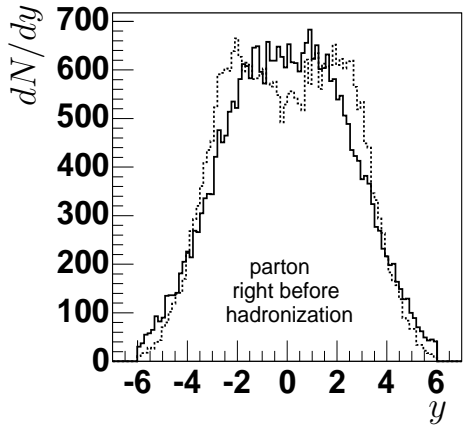}
\caption{\label{parton_rap} The rapidity distribution of partons
right before hadronization in AMPT v2.11 with parton-wise
hadronization (solid line) and with global phase transition
(dashed line.)}
\end{center}
\end{figure}

In order to see why the model with collective phase transition can
describe the experimental data better, the rapidity distributions
of partons right before hadronization are plotted in
Fig~\ref{parton_rap}. Comparing with the model with parton-wise
hadronization, the parton transport in the model with collective
phase transition is truncated, so there are fewer partons in the
mid-rapidity region and the distribution peaks at regions with
large absolute values of rapidity. This effect results in the
hadron distribution in mid-rapidity being lower than that from the
original model, and thus approaching the experimental data.

The elliptic flows in parton cascade models are built-up very
early~\cite{ampt_flow}~\cite{zpc_flow} (less than 5 fm/c), thus
the truncation of parton transport at 5 fm/c in the present work
does not affect the elliptic flow.

We have proposed a super-cooling followed by a global
hadronization as a prototype of the thermal-equilibrium phase
transition from parton transport to hadron gas. Though our method
for the implementation of phase transition seems to be very crude
comparing to the real process in relativistic heavy ion
collisions, the encouraging results indicate that a relevant
parton transport model for relativistic heavy ion collision should
take equilibrium phase transition into proper account.

\end{document}